\documentclass[%
 reprint,
 amsmath,amssymb,
 aps,
longbibliography,
]{revtex4-1}

\usepackage{graphicx}
\usepackage{dcolumn}
\usepackage{bm}

\begin{document}


\title{Metasurface engineering through bound states in the continuum}


\author{Anton~S.~Kupriianov$^{1}$}
\author{Yi Xu$^{2}$}
\author{Andrey Sayanskiy$^{3}$}
\author{Victor Dmitriev$^{4}$}
\author{Yuri~S.~Kivshar$^{3,5}$}
\author{Vladimir~R.~Tuz$^{6,7}$}
 \email{tvr@jlu.edu.cn; tvr@rian.kharkov.ua}
\affiliation{$^1$College of Physics, Jilin University, Changchun 130012, China}
\affiliation{$^2$Department of Electronic Engineering, College of Information Science and Technology, Jinan University, Guangzhou 510632, China}
\affiliation{$^3$ITMO University, St.~Petersburg 197101, Russia}
\affiliation{$^4$Electrical Engineering Department, Federal University of Para, PO Box 8619, Agencia UFPA, CEP 66075-900 Belem, Para, Brazil}
\affiliation{$^5$Nonlinear Physics Center, Australian National University, Canberra ACT 2601, Australia}
\affiliation{$^6$International Center of Future Science, State Key Laboratory of Integrated Optoelectronics, College of Electronic Science and Engineering, Jilin University, Changchun 130012, China}
\affiliation{$^7$Institute of Radio Astronomy, National Academy of Sciences of Ukraine, Kharkiv 61002, Ukraine}  

              
\begin{abstract}
Metasurfaces have attracted a lot of attention in recent years due to novel ways they provide for the efficient wavefront control and engineering of the resonant transmission. We discuss an approach allowing effectively control appearance of the sharp Fano resonances in metasurfaces associated with the {\it bound states in the continuum}. We demonstrate that by breaking the symmetry transversely, in the direction perpendicular to a metasurface with a complex unit cell, we can control the number, frequency, and type of high-$Q$ resonances originating from bound states in the continuum. As example, we demonstrate experimentally the metasurfaces with magnetic dipole and toroidal dipole responses governed by the physics of multipolar bound states.  
\end{abstract}

\maketitle
\section{\label{intr}Introduction}

Bound states in the continuum (BICs) are waves that remain localized in dynamic systems even though they coexist with a continuous spectrum of radiating waves. Although BICs were first introduced in quantum mechanics \cite{vonNeumann1929, Stillinger_PhysRevA_1975}, it is a general wave phenomenon that has been found in many physical systems, such as electromagnetic, acoustic, water and elastic waves \cite{ursell_1951, Robnik_1986, Ursell_London_1991, Evans_qjmam_1993}. Specifically, these states have also been identified in a wide range of optical systems, including dielectric gratings \cite{Shabanov_PhysRevLett_2008}, optical waveguides \cite{Shipman_PhysRevE_2005, Bulgakov_PhysRevB_2008, Segev_PhysRevLett_2011, Weimann_PhysRevLett.111.240403}, photonic crystals \cite{LeePhysRevLett.109.067401, hsu2013observation}, graphene quantum-dot structures \cite{Gonzalez_EPL_2010}, and hybrid plasmonic-photonic systems \cite{Shalaev_PhysRevLett.121.253901} (see a comprehensive review on BICs in Ref.~\onlinecite{Soljacic_NatRevMat_2016} and references therein). 

One interesting type of BICs relies on the symmetry properties of both discrete mode and coexisted radiative continuum. When a discrete mode is characterized by a particular symmetry class, radiative continuum belonging to a different symmetry classes completely decouples from such a discrete state \cite{LeePhysRevLett.109.067401, Miroshnichenko_PhysRevB_2018}. The coupling between the bound states and continuum band is forbidden as long as their symmetry properties are preserved, and they are classified as symmetry-protected BICs \cite{LeePhysRevLett.109.067401}.

A particular signature of symmetry-protected BICs in the field of optical metasurfaces is the observation of exceptionally narrow resonances (sometimes called {\it trapped modes}) in the optical spectra of metasurfaces composed of in-plane asymmetric structural elements \cite{Zouhdi_Advances_2003, Fedotov_PhysRevLett_2007}. The appearance of ultrahigh-$Q$ resonances are attributed to the excitation of asymmetric modes whose electromagnetic field distributions are slightly deviated from the, otherwise inaccessible, symmetric modes. The corresponding quality factors of leaky resonances depend on the degree of the introduced structural asymmetry \cite{Kivshar_PhysRevLett_2018}. The concept of manipulating the transition between the symmetry-protected BICs and leaky resonances was first proposed for plasmonic metasurfaces composed of metallic split-ring resonators \cite{khardikov_JOpt_2010}, and then it was developed for all-dielectric metasurfaces \cite{Khardikov_JOpt_2012, Zhang_OptExpress_2013, jain_advoptmater_2015, Khardikov2016, Kruk_ACSPhotonics_2017, tuz_ACSPhotonics_2018, Tuz_OptExpress_2018, Sayanskiy_PhysRevB_2019, tuz_JApplPhys_2019, tuz_AdvOptMat_2019}. 

\begin{figure}[tp]
\centering
\includegraphics[width=1.0\linewidth]{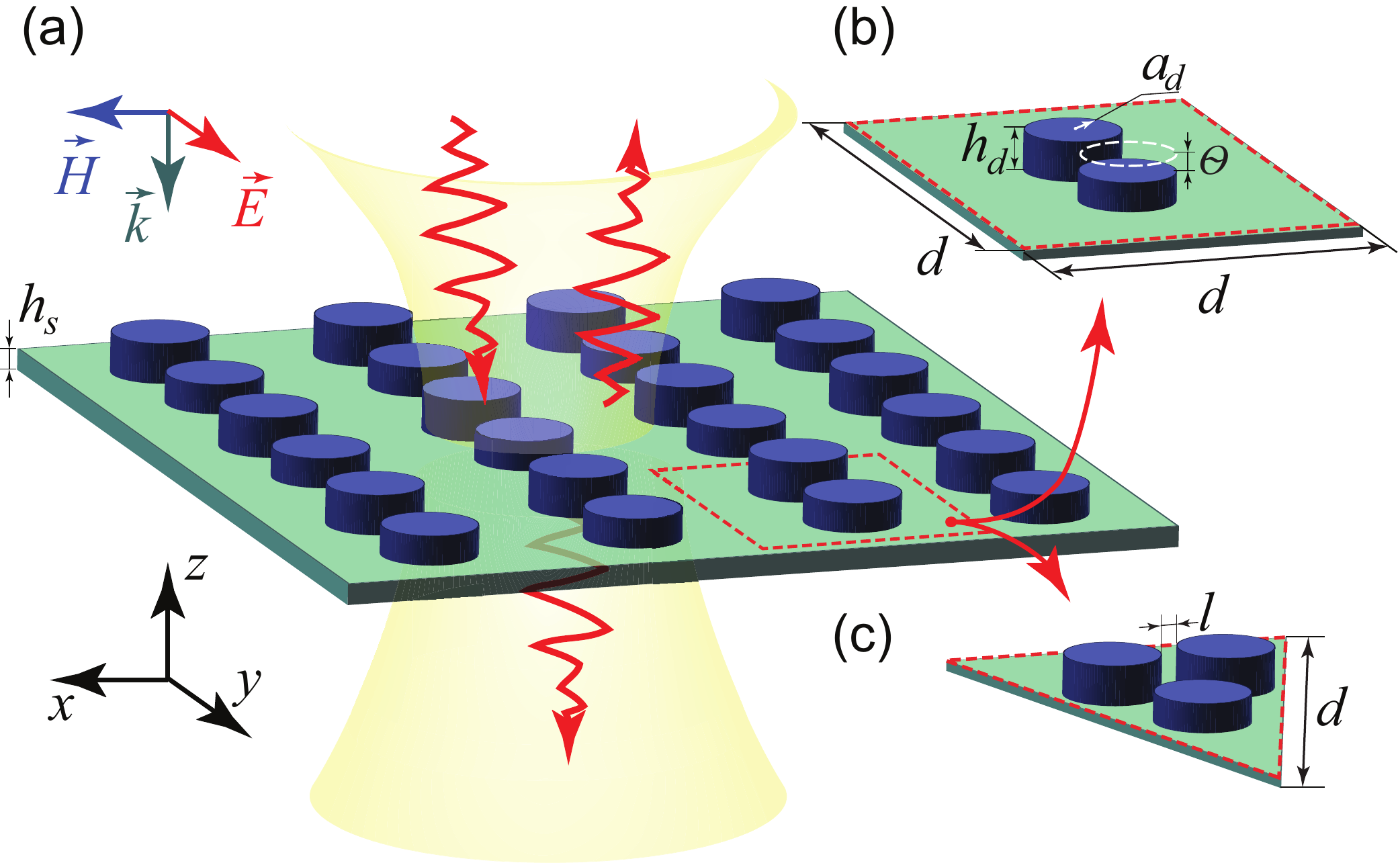}
\caption{(a) The symmetry breaking in all-dielectric metasurfaces composed of cluster-based unit cells; two typical examples are (b) dimer and (c) trimer clusters of dissimilar subwavelength dielectric disks.}
\label{fig:sketch}
\end{figure}

The recent development in topological photonics has demonstrated a wide class of topologically protected states \cite{soljacic_NatPhoton_2014}. It has been demonstrated that, depending on the characteristics of the topology of the bulk bands, topological edge modes can be protected to certain classes of defects \cite{Zhen_PhysRevLett_2014}. It is revealed \cite{Fleury_PhysRevLett_2019} that there is a correspondence between electromagnetic BICs and a wide range of topologically related phenomena including Berry phases around Dirac points  \cite{Manes_PhysRevB_2007}, topological defects \cite{Mermin_RevModPhys_1979}, and general vortex physics \cite{Donnelly_book_1991}. There are prospects in the field of topological photonics with the use of cluster-based all-dielectric metasurfaces \cite{wu_PhysRevLett_2015, Slobozhanyuk_NatPhoton_2016} mediated by BICs. 

In general, BIC modes transform into leaky resonances when the symmetry of either eigenmodes of the structure or radiative continuum is broken. As a result, one needs to break the symmetry of either discrete mode or radiative continuum in order to access such symmetry protected multipolar BICs. To date, all the mechanisms reported so far for manipulating the transition between BICs and radiative continuum can be classified as \textit{in-plane symmetry breaking}. No effort has been paid to manipulate the transition between BICs and leaky resonances utilizing the \textit{out-of-plane symmetry}, which will provide more flexible approach to engineer and control the resonant properties of metasurfaces.

In the present paper, we introduce an approach allowing effectively control the high-$Q$ Fano resonances, which are originated from the BICs of optical metasurfaces, by utilizing out-of-plane symmetry breaking in the unit cell of an all-dielectric metasurface. We demonstrate that our idea is quite general, which has been validated in metasurfaces composed of lattices of particle clusters corresponding to different symmetries. By introducing this unique degree of freedom beyond the in-plane symmetry, one can efficiently control the symmetry of the discrete mode and thus manipulate its interaction with the continuum. Employing this approach, we realize accesses to various kinds of multipolar BIC associated leaky modes existed in the cluster-based all-dielectric metasurfaces. We further perform proof-of-principle microwave experiments by demonstrating leaky resonances associated with the toroidal dipole BICs in both far-field and near-field measurements.

\section{\label{dimer} Toroidal quasi-BIC modes}

Without loss of generality, we consider an all-dielectric metasurface whose unit cell consists of different types of particle clusters [Fig.~\ref{fig:sketch}(a)]. The simplest configuration of the unit cell is a disk dimer arranged in a square lattice, as shown in [Fig.~\ref{fig:sketch}(b)]. Therefore, we start from the demonstration of our approach by evaluating how the out-of-plane symmetry breaking will affect on the transition between various BIC modes and the corresponding leaky resonances in a dimer-based metasurface. Then it is generalized to a more complex trimer-based metasurface composed of equilateral triangular unit cells [Fig.~\ref{fig:sketch}(c)]. 

All notations and parameters related to the geometry of the metasurface under study are summarized in Fig.~\ref{fig:sketch}. The disks are made of a non-magnetic material with relative permittivity $\varepsilon_d$ and the metasurface is placed on a dielectric substrate with relative permittivity $\varepsilon_s$. We consider that the metasurface is illuminated by a normally incident ($\vec k = \{0,0,-k_z\}$) linearly polarized plane wave with the electric field vector directed either along the $x$ axis ($\vec E = \{E_x,0,0\}$, $x$-polarized wave) or the $y$ axis ($\vec E = \{0,E_y,0\}$, $y$-polarized wave).

\begin{figure}[tp]
\centering
\includegraphics[width=1.0\linewidth]{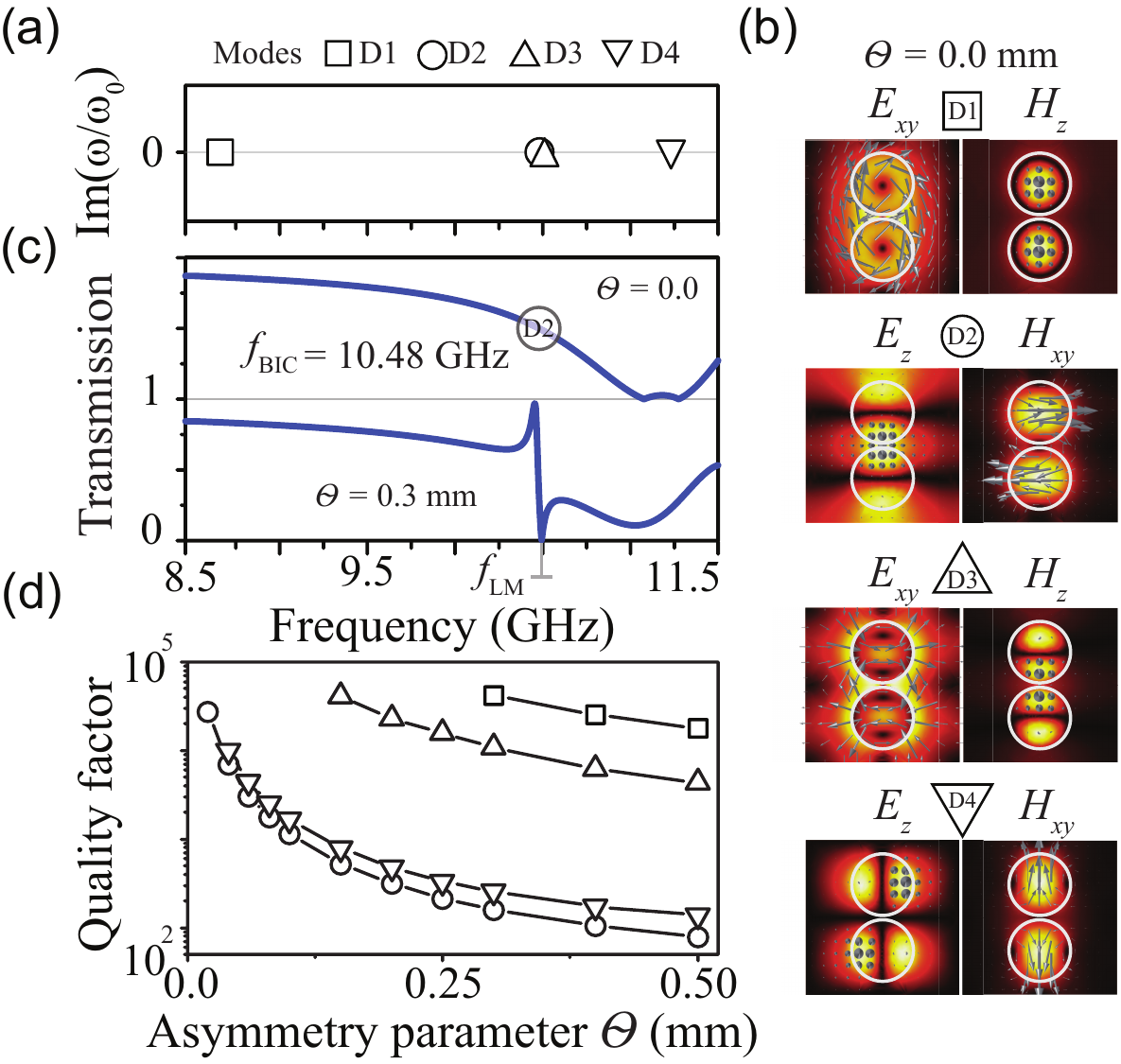}
\caption{(a) A set of the BIC-type eigenstates and (b) the corresponding electromagnetic near-field distribution for metasurfaces with the unit cell composed of a pair of disks (dimer). (c) Access to leaky resonance originated from the BIC Mode D2 of the dimer metasurface via out-of-plane symmetry breaking. (d) Dependence of the quality factors of leaky resonances originated from BIC modes D1-D4 on the out-of-plane asymmetry parameter $\theta$ (in [mm]). Parameters of the metasurface are: $r_d = 4$~mm, $h_d = 3.5$~mm, $d = 20$~mm, $l = 0.5$~mm, $\varepsilon_d = 22$, $\varepsilon_s = 1$.}
\label{fig:simulated_di}
\end{figure}

For the metasurface under study we can recognize the BIC modes by performing an eigenvalue analysis using the RF module of commercial COMSOL Multiphysics\textsuperscript{\textregistered} finite-element electromagnetic solver \cite{comsol}. We used this solver also to calculate electromagnetic patterns of the modes. The BIC modes can be distinguished from other ones by their real-valued eigenfrequencies. In accordance with our experimental possibilities, we consider the BIC modes in the microwave part of the spectrum ($8-12$ GHz). From the known set of modes we selected several BIC modes which appear in the chosen frequency band for the given parameters of the structure. We denoted the modes of interest by symbols D1-D4 [Fig.~\ref{fig:simulated_di}(a)]. These BIC modes manifest themselves by distinct displacement current geometries, which can be regarded as specific combinations of magnetic dipole (MD) and electric dipole (ED) moments inherent to an individual disk [Fig.~\ref{fig:simulated_di}(b)]. 

Without making a change in the material composition of the metasurface, accessing to the BIC modes can be realized by the symmetry breaking of unit cell with the help of modification of the particles' geometry inside the cluster. Such a modification can be done in either or both in-plane and out-of-plane directions of the metasurface, where distinct sensitivities can be expected referring to different symmetry protected BIC modes. The required change in the geometry can be determined from the analysis of the electromagnetic field pattern of the corresponding mode. For the BIC modes in which the MD moment lies in plane ($x$-$y$ plane), their turning to leaky modes can be effectively achieved by breaking the out-of-plane symmetry.

In particular, for Mode D2 the magnetic field vector is anti-parallel located in the $x$-$y$ plane, whereas displacement current flows inside the particles in the $y$-$z$ plane and has a solenoid-like behavior enclosing the magnetic field. It resembles the typical out-of-plane toroidal dipole (TD) BIC mode which was theoretically studied recently \cite{Miroshnichenko_PhysRevB_2018}. Rather than enabling the coupling between TD BIC mode and the obliquely incident plane wave, we show that this TD BIC mode can be turned to a TD leaky resonance by breaking the out-of-plane symmetry, as shown in Fig.~\ref{fig:simulated_di}(c) (we used the Comsol Multiphysics solver to calculate the transmission characteristic of the metasurface as well). Specifically, such a coupling to Mode D2 can be realized for the $y$-polarized incident wave as soon as a difference in the thicknesses of disks forming a dimer is introduced [we denote this difference as the asymmetry parameter $\theta$ introduced in the units of length, see Fig.~\ref{fig:sketch}(b)]. Notice that the group of symmetry in the $x$-$y$ plane of the non-perturbed  dimer with $\theta=0$ is $C_{2v}$, and of the perturbed dimer with $\theta \neq 0$ is $C_{s}$.

Indeed, while the symmetry of dimer is preserved, there is no peculiarities in the transmitted spectra of the metasurface under study at the frequencies of the BIC modes [the eigenfrequency of Mode D2 is denoted as $f_\textrm{BIC}$, see Fig.~\ref{fig:simulated_di}(c)]. As soon as the asymmetry is introduced, there is a leaky resonance (the resonant frequency of the leaky mode originated from Mode D2 is denoted as $f_\textrm{LM}$) in the transmission spectrum and such resonance acquires a peak-and-trough (Fano) profile which is typical for the metasurfaces composed of in-plane asymmetric structural elements \cite{Fedotov_PhysRevLett_2007}. The Fano resonance is highly asymmetric, in which the asymmetry depends on the parameter $\theta$. 

The $Q$-factor of the Fano resonance is further defined as $Q=f_\textrm{LM}/\Delta f$, where $\Delta f$ is the full width half maximum of the mode \cite{Zhang_NanoLett_2011}. By controlling the asymmetry factor, we can manipulate the $Q$-factors of the leaky resonances. The less out-of-plane asymmetry is, the higher the quality factor of the leaky mode resonance is [Fig.~\ref{fig:simulated_di}(d)]. At the same time, the corresponding frequencies $f_\textrm{BIC}$ and $f_\textrm{LM}$ related to Mode D2 are found to be different but they are rather closely spaced to each other. The difference between them also depends on the degree of the out-of-plane symmetry breaking. Therefore, breaking the out-of-plane symmetry of the unit cell in the metasurface presents an efficient way to manipulate the quality factor of resonances related to the BIC modes. 

It should be pointed out that breaking the out-of-plane symmetry of the cluster composed of disks has fundamentally different effects on the quality factor of the BIC modes compared with the case of breaking the in-plane symmetry. One can anticipate that introduced asymmetry in the thicknesses of disks forming the dimer has a stronger effect on its out-of-plane ED moment and in-plane MD moment, since the polarization currents are parallel to such plane. In contrast, breaking the in-plane symmetry by changing the radius of one disk has more pronounced effect on the in-plane ED moment and out-of-plane MD moment. For instance, among the selected BIC modes of dimer, Mode D2 is considered as a TD BIC composed of two anti-parallel in-plane MD moments. Therefore, symmetry breaking in the out-of-plane direction turns such a TD-BIC into a leaky resonance much more efficiently compared with other types of BIC modes. At the same time, the decrease of $Q$-factor for Mode D4 consisted in the in-plane MD moments is also faster than that for Modes D1 and D3, as shown in Fig.~\ref{fig:simulated_di}(d). As a result, the out-of-plane symmetry breaking can be regarded as a new degree of freedom and an efficient way to tailor the resonant properties of the symmetry protected BIC modes, especially for the out-of-plane TD BIC and in-plane magnetic BIC modes.

\section{\label{trimer}Quasi-BIC mode manipulation}

We further present an example utilizing the out-of-plane symmetry breaking induced transition between the symmetry protected BIC modes and the corresponding leaky resonances in an all-dielectric metasurface based on trimers [Fig.~\ref{fig:sketch}(c)]. The existence of the TD mode in the metasurface whose lattice is formed by trimers arranged in square unit cells was examined in \cite{tuz_AdvOptMat_2019}. Here we introduce some modifications to the structure of unit cell studied earlier to provide further degree of freedom for resonance manipulation in both near- and far-fields. Firstly, we consider that the unit cell has the form of an equilateral triangle. Secondly, two triangle unit cells are paired together in such a way that a rhomboid cluster appears, i.e., generally, the metasurface under study is constructed from the twin-trimer super-cells.

\begin{figure*}[htp]
\centering
\includegraphics[width=1\linewidth]{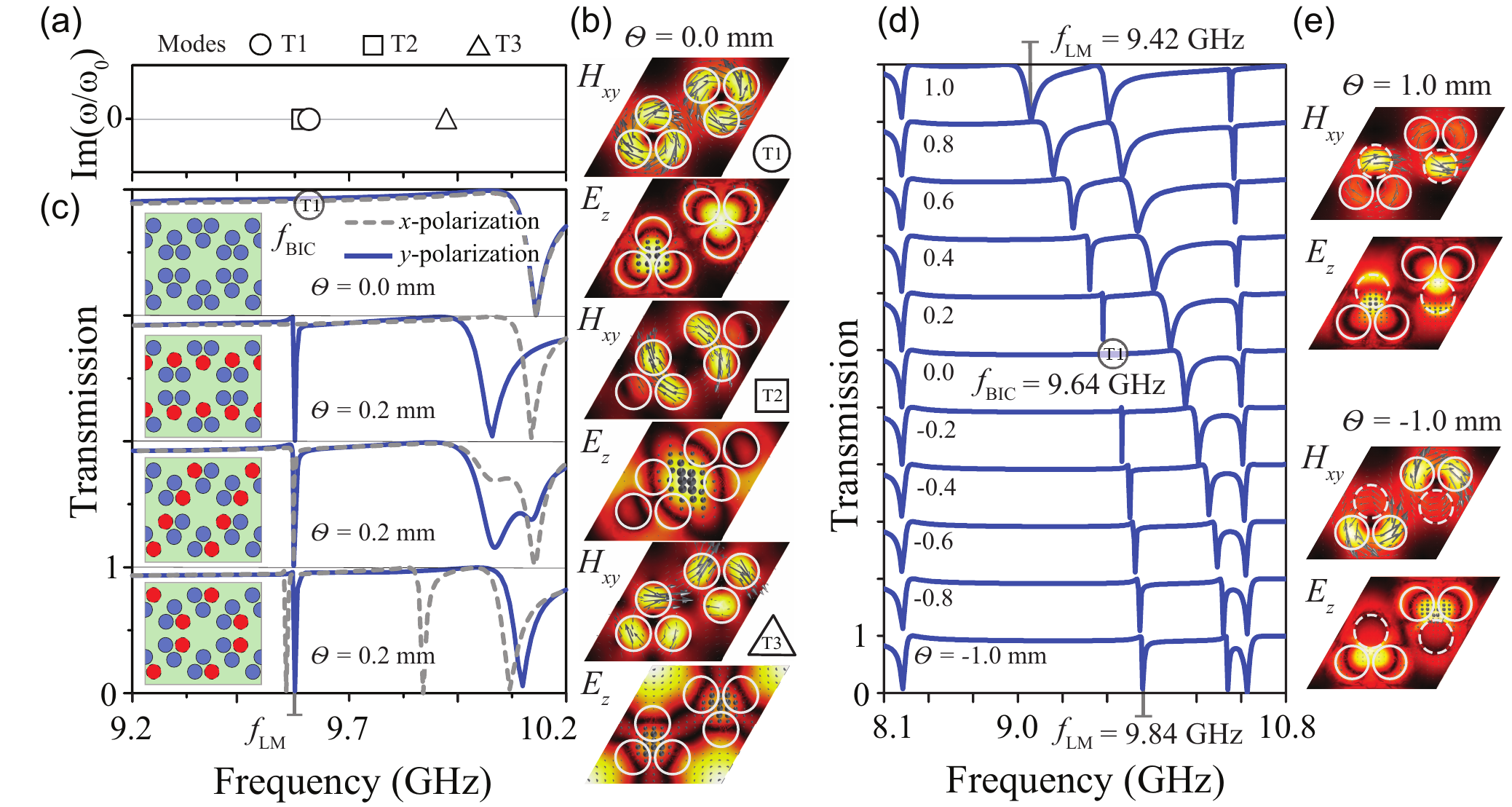}
\caption{(a) A set of BIC-type eigenstates and (b) their electromagnetic near-field distributions for metasurfaces with the  twin-trimer unit cell. (c) Schemes of access to Mode T1 in the trimer-based metasurface. (d) Evolution of the transmission spectra vs the asymmetry parameter $\theta$ (in [mm]). (e) Electromagnetic near-field distribution of the leaky mode at the resonant frequency corresponding to the manifestation of Mode T1. $d = 31.1$~mm, all other parameters are the same as in Fig.~\ref{fig:simulated_di}.}
\label{fig:simulated_tr}
\end{figure*}

After the eigenvalue analysis of this metasurface with the COMSOL Multiphysics electromagnetic solver, a set of BIC modes existed in the frequency band of interest are selected. We denote these modes by symbols T1-T3 [Fig.~\ref{fig:simulated_tr}(a)]. All these modes present fruitful electromagnetic near-field distributions manifested as electric and magnetic hot-spots in the near-field [Fig.~\ref{fig:simulated_tr}(b)]. Specifically, these BIC modes are composed of various MD moments in each disk, which are arranged in the $x$-$y$ plane. When all disks are identical, the symmetry of the twin-trimer super-cell is $C_{2v}$ and it belongs to the rhomboid system (see Fig.~\ref{fig:simulated_tr}(c), upper part). As can be seen from this figure, neither the $x$-polarized nor the $y$-polarized normally incident plane wave can access to these BIC modes. As the in-plane MD moment is more sensitive to the out-of-plane asymmetry for the all-dielectric metasurface under consideration, all these BIC modes can be effectively turned into leaky resonance by increasing the thickness of certain disks. For example, perturbation of the unit cell by two disks with different thicknesses reduces the symmetry to $C_2$ of the monoclinic system (Fig.~\ref{fig:simulated_tr}(c), middle part). This perturbation allows one to excite the MD leaky resonance originated from BIC Mode T1. It should be pointed out that although this metasurface has a polarization-dependent response, the corresponding resonance may be excited by the $x$-polarized as well as the $y$-polarized plane waves [Fig.~\ref{fig:simulated_tr}(c)] (a polarization-independent design should be seek in the metasurfaces composed of hexagonal unit cells considering symmetries of the group C$_6$ \cite{overvig2019selection}).

At the same time, the out-of-plane symmetry breaking can also be used to tailor the frequency and $Q$-factor of the leaky Fano resonance. By changing the asymmetry factor $\theta$ from $1$~mm to $-1$~mm, one can observe the shift of the resonant frequency of the leaky mode $f_\textrm{LM}$ compared with the eigenfrequency $f_\textrm{T1}$ of Mode T1, as shown by Figs.~\ref{fig:simulated_tr}(d). The electromagnetic field distributions at the resonant frequency under the excitation of the $y$-polarized plane wave are shown in Figs.~\ref{fig:simulated_tr}(e) for two typical asymmetry parameters $\theta = 1$~mm and $\theta = -1$~mm, respectively. It can be observed that the $Q$-factor of this resonance can also be manipulated accordingly by using different degrees of out-of-plane symmetry breaking. More importantly, the electromagnetic near-field can also be manipulated by using the asymmetry parameter $\theta$.

It is noticed that the leaky mode $f_\textrm{LM}$ appears to be robust when $\theta$ is negative [Figs.~\ref{fig:simulated_tr}(d)], thus showing a weak renormalization of the BIC frequency when converted to a quasi-BIC as a result of the symmetry-breaking. This effect is due to the difference in the electromagnetic field localization and distribution in the particles forming the cluster. From Figs.~\ref{fig:simulated_tr}(e) it can be seen that for $\theta  = 1$~mm the field is largely localized in perturbed particles, where perturbation in that particle results in more sensitivity in the resonant frequency of the quasi-BIC mode, whereas when $\theta  = -1$~mm it is in the non-perturbed ones. 

Therefore, varying parameter $\theta$ can result in different electromagnetic near-field distributions though the corresponding leaky mode is originated from the same BIC Mode T1, which validates the ability to manipulate the vector near fields by out-of-plane symmetry breaking. It should be pointed out that our approach is quite general and it can be easily generalized to resonant metasurfaces operating at the optical spectrum. The ability to control the electromagnetic near-field by utilizing out-of-plane symmetry breaking provides possibility to precisely tailor both the emission of the ED and MD types of dipole emitters \cite{Feng_2018}. 

\section{\label{experiment}Experimental demonstrations}

To experimentally verify the out-of-plane symmetry breaking effect as an effective mean to trigger the transition between BIC modes and high-$Q$ Fano resonance, an all-dielectric metasurface is fabricated and studied in the microwave range. A prototype of the twin-trimer metasurface was prepared. Two sets of dielectric disks were made from Taizhou Wangling TP-series polymer-ceramic composite, whose permittivity is $\varepsilon_d = 21$ and loss tangent is $\tan\delta_d\approx 1\times 10^3$ at $10$~GHz. The dielectric particles have the same diameter of $8$~mm, while their thicknesses are $3.5$~mm and $2.5$~mm, respectively. To arrange them into a lattice, an array of holes was milled in a custom holder made of a Styrofoam material whose relative permittivity is $1.03$. The thickness of the holder is $20$~mm and the depth of the hole is $3.5$~mm. The prototype of metasurface is composed of $15 \times 26$ trimers in total. 

At the first step the transmission spectra have been measured. The common technique when the measurements are performed in the radiating near-field region and then transformed to the far-field zone has been used \cite{johnson_IEEE_1973}. During the investigation the prototype was fixed on the $2.5$~m distance from a circular linearly polarized horn antenna working in the X frequency range ($8-12$~GHz). The antenna generates a quasi-plane-wave with required polarization. The antenna is connected to the port of the Keysight E5071C Vector Network Analyzer (VNA) by a $50$~Ohm coaxial cable. An electrically small dipole probe oriented in parallel to the metasurface plane and connected to the second port of the VNA was used to detect the corresponding electric field component. The sketch and photo of our experimental setup can be found in Refs.~\onlinecite{Sayanskiy_PhysRevB_2019} and \onlinecite{tuz_AdvOptMat_2019}. 

\begin{figure}[!t]
\centering
\includegraphics[width=1.0\linewidth]{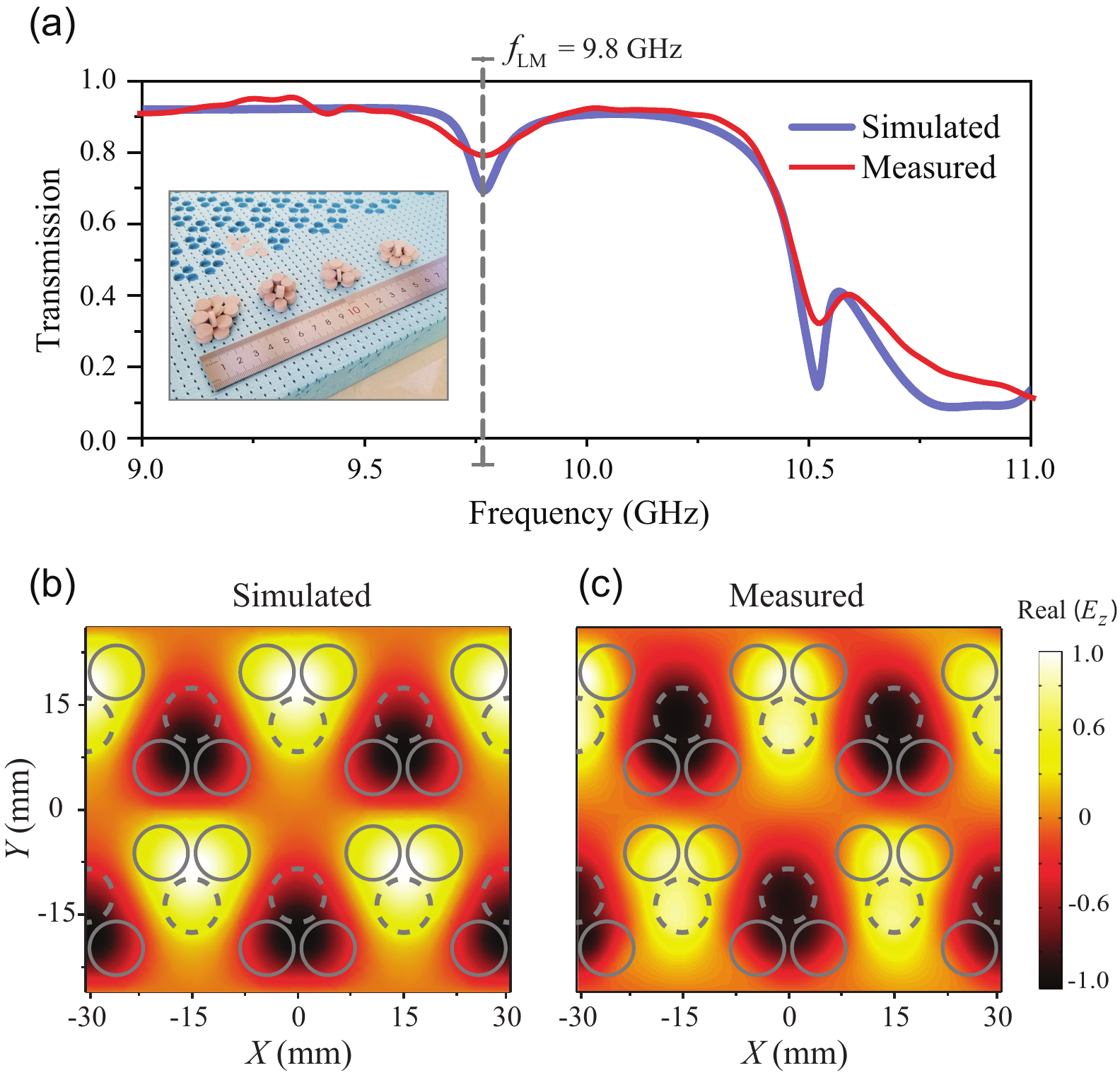}
\caption{Experimental results. (a) Transmission coefficient of the trimer-based metasurface. The inset presents a fragment of the metasurface prototype and several sets of particles. In the simulation actual material losses ($\tan\delta=1 \times 10^{-3}$) in ceramic particles are taken into account, while the substrate is modeled as a loss-less dielectric. The asymmetry factor is $\theta=-1.0$~mm. All other material and geometrical parameters of the metasurface are the same as in Fig.~\ref{fig:simulated_tr}. (b) Simulated and (c) measured near-field distribution of the real part of the normal component of the electric field ($E_z$) at the resonant frequency $f_\textrm{LM}$. The measurement plane is 1 mm away from the metasurface. The field is normalized to the maximum value of $E_z$.}
\label{fig:measured_tr}
\end{figure} 

To scan a certain area at the $x$-$y$ plane, a Linbou NFS03-type three-dimensional automatic scanning system was used. The scan area ($200 \times 200$~mm) was chosen a little smaller than the sample size in order to reduce the effects of the sample edges on the measurement results. During the measurements the probe was automatically moved in the $x$-$y$ plane over the scan area with a $5$~mm step at the distance of $50$~mm above the prototype surface. At each probe position both amplitude and phase of the transverse component of the electric field were sampled in the frequency range of interest. The same value of the electric field in the absence of the prototype was measured as a reference and the post-processing procedure was performed to obtain the transmission coefficient \cite{johnson_IEEE_1973}. To reduce the undesired reverberations between the horn and the metasurface, the time-gating technique is applied. 

For the near-field mapping measurement of longitudinal component of the electric field at the frequency $9.77$~GHz, a Linbou NFS03-type three-dimensional automatic scanning system was adopted to observe the electromagnetic near-field distribution directly at the distance of $1$~mm above the prototype surface. The same horn antenna is connected to the port $1$ of the VNA and a monopole antenna perpendicularly oriented to the sample connected to the second port. The probe was automatically moved in the $x$-$y$ plane over the scan area $94 \times 54$ with a $1$~mm step at the distance of $1$~mm above the prototype surface. At each probe position both amplitude and phase of the longitudinal component of the electric field were measured.

The manifestation of the corresponding leaky resonance is detected in the far-field measurement of the transmission coefficient of the metasurface [Fig.~\ref{fig:measured_tr}(a)]. In order to provide a reference for experiment, we present the simulated electric near-field distribution [Real$(E_z)$] of the metasurface at the frequency $f_\textrm{LM}$. As can be seen from Fig.~\ref{fig:measured_tr}(b), the electric field is mostly concentrated in the center of trimers, being out of phase in the pair of trimers forming the super-cell. The electromagnetic near field distribution are then experimentally measured for the all-dielectric metasurface at the same resonant frequency $f_\textrm{LM}$. Indeed, the simulated [Fig.~\ref{fig:measured_tr}(b)] and experimental [Fig.~\ref{fig:measured_tr}(c)] results show excellent agreement with each other, validating the excitation of MD leaky resonance originated from BIC Mode T1. 

\begin{figure}[!t]
\centering
\includegraphics[width=1.0\linewidth]{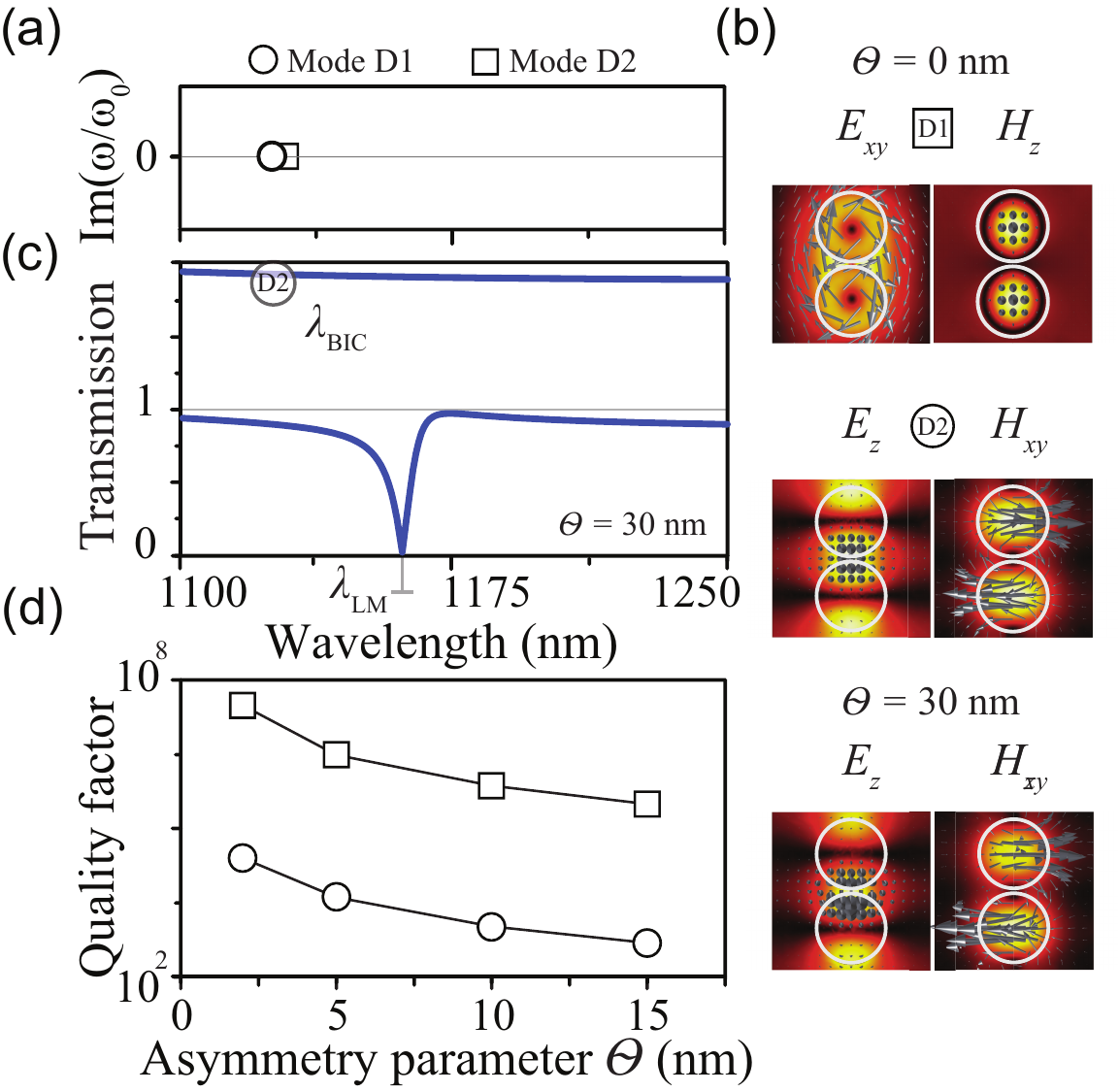}
\caption{Metasurface with a dimer cluster in a square unit cell. (a) A set of eigenstates behaving as BICs and (b) their electromagnetic near-field distribution for the dimer unit cell. (c) Schemes of access to Mode D2 in the dimer-based metasurface. The transmission spectra of the metasurface with non-perturbed and perturbed clusters. (d) Dependence of the quality factors of leaky resonances originated from BIC modes D1 and D2 on the out-of-plane asymmetry parameter $\theta$ (in [nm]). The structure is constructed from silicon nanodisks ($\varepsilon_d=11.56$) separated by a gap $l=30$~nm. The thickness of nanodisks is $h_d=220$~nm and the radius is $a_d=160$~nm. The period of metasurface is $p=730$~nm.}
\label{fig:optic_dimer}
\end{figure} 

\section{\label{app} Concluding remarks}

Importantly, our approach for engineering metasurfaces with BIC is rather general, and it can be readily applied to silicon metasurfaces with BIC-driven resonant frequencies in the optical spectral range ($1100-1250$~nm). To demonstrate that, we show in Fig.~\ref{fig:optic_dimer} one example of optical metasurfaces based on silicon dimer clusters arranged in a square unit cell. Transmission spectra of the metasurface with non-perturbed and perturbed clusters clearly show how the $Q$-factors of leaky resonances originate from BIC modes defined by the out-of-plane asymmetry parameter $\theta$. One can see in Fig.~\ref{fig:optic_dimer}(d) the $Q$-factor of mode D2 in the optical metasurface ($Q\sim 10^7$) is sufficiently higher than that one obtained in the microwave prototype ($Q\sim 10^4$) [Fig.~\ref{fig:simulated_di}(d)]. Compared with actual material losses in ceramic particles used in our microwave experiments, the losses in silicon particles in the chosen wavelength range are negligible \cite{palik1997handbook}, so it is reasonable to expect the quality factor of the quasi-BIC resonance in silicon-based metasurfaces can reach several thousands in a finite structure. This state with a high quality factor provides strong field confinement near the surface, making it potentially useful for chemical or biological sensing, organic light emitting devices and large-area laser applications. 

In summary, we have discussed a general approach for a design of highly efficient metasurfaces with controlled high-$Q$ resonances via the symmetry breaking in two dimensional planar lattices in the out-of-plane direction, which essentially trigger the transition between symmetry-protected BIC modes and leaky modes with quasi-BIC resonances. We have demonstrated that both the resonance frequency and the value of the $Q$-factor of the leaky resonances can be manipulated by changing the asymmetry parameter $\theta$. At the same time, the electromagnetic near-field structure can also be tailored simultaneously to address the electric and magnetic hot-spots locally. As an example, we have presented microwave experiments to validate our general concept of the out-of-plane symmetry breaking for the efficient metasurface engineering and their $Q$-factor control. 

Using trimer-based hexagonal unit cells with the symmetry of the group C$_6$ or C$_{6v}$, one can expect the polarization-independent response of such a  metasurface. Verification of this statement, however, requires further investigation. 

In the present paper we restrict ourselves to the metasurfaces illuminated by linearly polarized waves. Therefore, excitation of the quasi-BIC modes by waves of another polarization or by structured light could be another degree of freedom to tailor the quasi-BIC modes, which however goes beyond the scope of our investigation and calls for further study.

Finally, we would like to notice that, because both the Fano resonance and bound state in the continuum are general wave phenomena, the similar asymmetry-driven quasi-BIC resonances are expected to occur for other types of waves, so we may to find them in acoustic metamaterials or elastic waves.

\begin{acknowledgements}
A.S. and V.R.T. acknowledge the Jilin University for hospitality and financial support; V.D. thanks the Brazilian agency CNPq for a financial support; Y.S.K. acknowledges a support from the Australian Research Council and the Strategic Fund of the Australian National University. 
\end{acknowledgements}

\bibliography{magnetic_bic}

\end{document}